# *Cell death induced by the application of alternating magnetic fields to nanoparticle-loaded dendritic cells.*

I. Marcos-Campos[1,2], L. Asín[1], T. E. Torres[1,3,4], C. Marquina[3,4], A. Tres[1,2], M. R. Ibarra[1,3,4] and G. F. Goya[1,3,†],

[1] *Instituto de Nanociencia de Aragón (INA), University of Zaragoza, Pedro Cerbuna 12, 50009- Zaragoza, Spain.*
[2] *Oncology Department, Hospital Universitario "Lozano Blesa", 50009 Zaragoza, Spain.*
[3] *Condensed Matter Department, Sciences Faculty, University of Zaragoza, Spain 50009.*
[4] *Instituto de Ciencia de Materiales de Aragón (ICMA), CSIC-Universidad de Zaragoza, 50009 Zaragoza, Spain*

[†] *Corresponding author: goya@unizar.es.*
*Instituto de Nanociencia de Aragón (INA), University of Zaragoza, Pedro Cerbuna 12,*
*50009- Zaragoza, Spain.*
*Tel.: +34 976 762777*
*Fac.: +34 976 762776.*

*Abstract.*

In this work, the capability of primary, monocyte-derived dendritic cells (DCs) to uptake iron oxide magnetic nanoparticles (MNPs) is assessed and a strategy to induce selective cell death in these MNP-loaded DCs using external alternating magnetic fields (AMFs) is reported. No significant decrease in the cell viability of MNP-loaded DCs, compared to the control samples, was observed after five days of culture. The amount of MNPs incorporated into the cytoplasm was measured by magnetometry, which confirmed that 1 to 5 pg of the particles were uploaded per cell. The intracellular distribution of these MNPs, assessed by transmission electron microscopy, was found to be primarily inside the endosomic structures. These cells were then subjected to an AMF for 30 min, and the viability of the blank DCs (i.e., without MNPs), which were used as control samples, remained essentially unaffected. However, a remarkable decrease of viability from approximately 90% to 2-5% of DCs previously loaded with MNPs was observed after the same 30 min exposure to an AMF. The same results were obtained using MNPs having either positive ($NH_2^+$) or negative ($COOH^-$) surface functional groups. In spite of the massive cell death induced by application of AMF to MNP-loaded DCs, the amount of incorporated magnetic particles did not raise the temperature of the cell culture. Clear morphological changes at the cell structure after magnetic field application were observed using scanning electron microscopy. Therefore, local damage produced by the MNPs could be the main mechanism for the selective cell death of MNP-loaded DCs under an AMF. Based on the ability of these cells to evade the reticuloendothelial system, these complexes combined with an AMF should be considered as a potentially powerful tool for tumour therapy.

## 1. Introduction

The increasing synergy between biomedicine and nanoscience has been applied to the use of magnetic nanoparticles (MNPs) in routine laboratory and clinical protocols, such as cell sorting, DNA separation, magnetic resonance imaging (MRI) and gene therapy.[1, 2] Applications of MNPs currently in preclinical stages include the cell-targeted delivery of anticancer agents and molecular diagnosis. [3, 4] Recently, magnetic hyperthermia strategies passed preclinical trial stages and received regulatory approval as a clinical protocol for thermotherapy.[5] Thus, with proper control of size and surface functionalisation, magnetic nanoparticles could become multifunctional intracellular agents for MRI, drug carriers and/or heat generators for hyperthermia. [6]

Loading MNPs with anticancer drugs and targeting to specific tumour sites is a promising strategy for the detection and elimination of neoplastic tissue, even at the single-cell level. However, a major obstacle for this approach is that the reticuloendothelial system (RES) detects and phagocytoses MNPs, preventing their targeting and therapeutic action.[7] Attempts to overcome this difficulty (i.e., to evade the RES) have been made through the functionalisation of MNPs to mimic biologic entities already present in the blood system, resulting in the generation of 'stealth' carriers.[8]

Eukaryotic cells can easily be 'targeted' *in vitro* with MNPs of different sizes. Different mechanisms have been reported for the detection and/or uptake of inorganic nanoparticles by cells, depending on the type of core material and surface coating of the particles used. The general consensus is that the mechanisms involved in particle incorporation are strongly dependent on the cell type.[9]

Dendritic cells (DCs) are the most important antigen-presenting cells, as they have a key role in the first steps of most immune responses. [10] Therefore, DCs are good candidates as therapeutic tools for immune diseases and malignant neoplasms. DCs are derived from bone marrow precursors and migrate to non-lymphoid tissues where they differentiate into dendritic cells.[11] As antigen-presenting cells, DCs are active at the membrane level and are thus good candidates to incorporate MNPs. Indeed, previously published reports have demonstrated that DCs can incorporate peptides, viral RNA,[12] bacterial DNA [13] and other molecules.[14] The uptake of antigens by DCs may occur by different processes, such as macropinocytosis, phagocytosis or receptor-mediated endocytosis [15]; more recently, the ability of DCs to incorporate several types of solid particles within a broad size range has also been described [16].

In the present work, the first results for a new strategy are demonstrated, based on the use of dendritic cells as natural carriers of magnetic nanoparticles. This investigation had two major goals: a) to study the inclusion and toxicity of magnetic nanoparticles in dendritic cells and b) to induce cell death by applying a time-varying magnetic field to the MNP-loaded DCs. The reason for using DCs as carriers of MNPs was to mimic biological units and to elude the immune response of the body. The advantage of using this strategy is that the cargo of the DCs could be designed to be either 'nude' magnetic MNPs for hyperthermia therapy or functionalised MNPs with specific drugs that could be released by the application of alternating magnetic fields (AMFs).

## *2. Materials and methods*

2.1. Magnetic nanoparticles

A set of 18 different types of commercially available magnetic colloids (*Micromod GmbH, Germany*) [17] having different hydrodynamic sizes and surface functional groups was tested with respect to absorption efficiency (see Figure S1 of supporting information). All of the evaluated MNPs were composed of a magnetite core ($Fe_3O_4$) coated with a properly functionalised dextran shell at the surface. The average hydrodynamic size of these colloids spanned from 20 to 500 nm. The purpose of this serial evaluation of the specific power absorption (SPA) in different colloids was to identify samples with maximum heating efficiency because the final goal was to induce cell death by the application of an AMF. The two samples with the highest SPA values and similar hydrodynamic radii but opposite surface charges (the filled bars in Figure S1 of supporting material) were selected. The criterion for having two sets of MNPs, one positively ($NH_2^+$) and one negatively ($COOH^-$) charged, was to avoid the possibility of decreasing the uptake affinity of the cells due to the surface charge of the MNPs. Samples with positively charged MNPs (positive because of amine groups at the surface, $NH_2^+$) and negatively charged MNPs (negative because of carboxyl groups at the surface, $COOH^-$) were labelled as NH2(MNP)+ and COOH(MNP)-, respectively. In addition to the specifications from the provider [18], the nanoparticles were characterised through transmission electron microscopy (TEM), dynamic light scattering (DLS), and magnetisation measurements.

The magnetic characterisation of colloids and cell cultures was performed in a commercial SQUID magnetometer (Quantum Design MPMS-XL) through static measurements as a function of field and temperature. Pure colloids were measured *as* *is* (i.e., in the liquid phase) after being conditioned in sealed sample holders with 200-µL capacity. Magnetisation data were collected in applied magnetic fields up to 5 T at temperatures between 5 K and 250 K to avoid the melting of the frozen liquid carrier. To determine the magnetic behaviour at room temperature, hysteresis loops M(H) were performed at T= 295 K in samples previously dried at room temperature. The same protocol was used for magnetic measurements in DC cultures.

## 2.2. Dendritic cell differentiation from PBMCs

Peripheral blood mononuclear cells (PBMCs) were isolated from normal blood by a density gradient (*Ficoll Histopaque-1077, Sigma*). Cells were washed twice with PBS and then centrifuged for 10 minutes at 800 rpm to avoid platelet contamination. The isolation of CD14+ cells was performed with magnetic beads (*CD14 Microbeads, Miltenyi*) by positive immunoselection using the autoMACS Separator (Miltenyi). CD14+ cells ($10^6$ cells/ml) were cultured in RPMI 1640 (*Sigma*) with 10% FBS, 1% glutamine, 1% antibiotics, IL-4 (25 ng/ml) and GM-CSF (25 ng/ml)(*Bionova*) for 5 days at 37ºC. Every two days, the medium was replaced by fresh medium containing the same concentration of interleukins. The phenotype for the DCs was characterised by staining with fluorochrome-labelled antibodies against CD45, CD14 (BD Biosciences, San Jose, CA), CD11c (eBioscience, San Diego, CA), CD40, CD83 and CD86 (Invitrogen, Carlsbad, CA). The flow cytometry analysis was performed with the FACSCalibur cytometer and the data were analysed with FACSDiva software.

## 2.3. Cytotoxicity assay

On the fifth day of the differentiation process, loosely adherent and non-adherent cells were collected, washed twice and seeded onto a 6-well plate at $10^6$ cells/ml (1 to $3 \times 10^6$ total cells per well). The cytotoxicities of COOH(MNP)- and NH2(MNP)+ colloids were evaluated by adding 50 μg MNP/ml medium into three wells each; the cell viability was measured by Trypan blue, MTT (3-(4,5-dimethylthiazol-2-yl)-2,5-diphenyltetrazolium bromide) and flow cytometry using annexin-propidium iodide markers on days 1, 3 and 5 after the addition of the MNPs. The Trypan blue assay was conducted by diluting 10 μl of cell samples into trypan blue (1:1).

For the MTT cell viability assay, $5 \times 10^4$ DCs taken from each sample were incubated with 20 μL (5 mg/ml) MTT for two hours at 37ºC, and the resultant formazan crystals were dissolved in 200 μL dimethyl sulfoxide. To avoid possible interference from remaining nanoparticles and/or cellular debris, all samples were previously spun down. The absorbance was measured by an microplate reader (ELISA) at 550 nm by placing 100 μl of each sample in a 96-well plate. The percent cell viability was calculated as: (absorbance treated cells/absorbance untreated cells) x100. Each assay was repeated three times.

DC viability was further checked by flow cytometry using a commercial kit (Immunostep, Spain). DCs ($10^6$) of each sample were resuspended in Annexin-binding buffer and stained with 5 μl of Annexin and 5 μl of propidium iodide. DCs were incubated for 15 minutes in the dark at room temperature. Analysis of the results was performed using a FACSAria Cytometer and FACSDiva Software.

2.4. Transmission electron microscopy (TEM)

Transmission electron microscopy (TEM) images were taken on an 80 kV TEM JEOL 1010. Samples were prepared by harvesting DCs on day 5 of differentiation, seeding them at $3 \times 10^6$ cells/well and incubating them with MNPs at a concentration or 50 μg MNP/ml medium. DCs without MNPs were used as negative controls. Fixation of the cells was performed using 4% glutaraldehyde in 0.1 M sodium cacodylate for 2 hours at 4ºC; the cells were then washed three times with PBS and subsequently fixed for 1 hour using potassium ferrocyanide (2.5%) in 0.1 M sodium cacodylate and 2% $OsO_4$. Finally, the cells were washed and dehydrated with acetone. The encapsulation was performed in EPON resin, and ultrathin sections (60-80 nm) were cut by ultramicrotomy; the sections were then mounted on a 400-mesh nickel grid. The deposited slides were further dried and treated with 2% uranyl acetate.

## 2.5. Specific power absorption experiments

The specific power absorption (SPA) of the pure colloids was measured with a commercial ac field applicator (model DM100 from nB nanoscale Biomagnetics, Spain) at $f = 260$ kHz and field amplitudes from 0 to 16 mT; the applicator was equipped with an adiabatic sample space (~ 0.5 ml) for measurements in the liquid phase. For *in vitro* magnetic hyperthermia experiments, DCs cultured as described above were collected and seeded onto a 6-well plate at $10^6$ cells/ml. COOH(MNP)- and NH2(MNP)+ nanoparticle suspensions (1% wt) were each added into two wells at a concentration of 50 μg MNP/ml medium. Two wells, each containing $10^6$ cells/ml without nanoparticles, were used as a control. Cells were cultured overnight at 37ºC. The next day, cells were collected from each well and washed four times with fresh medium to remove the MNPs that were not incorporated. Cells were counted and resuspended in 500 μl of the medium.

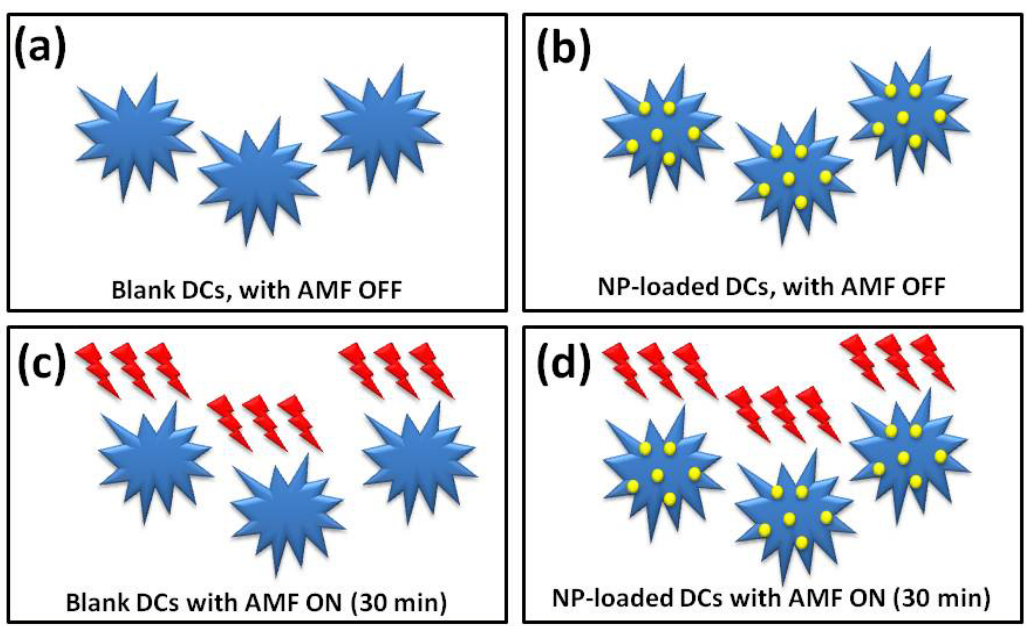

**Figure 1:** *Schematic representation of the '2x2' experiment: (a) DCs with neither MNP uptake nor magnetic field application. (b) DCs uploaded with MNPs but without field application, (c) DCs without MNPs for magnetic field application, and (d) DCs with MNPs and the application of magnetic fields.*

Each experiment consisted of four samples of cultured DCs, as illustrated in Figure 1. In this type of '2x2 experiment', the first pair of samples, which consisted of *as-cultured* blank DCs (i.e., without nanoparticles, Fig. 2a) and MNP-loaded DCs (Fig. 2b), were not exposed to magnetic fields and analysed at the end of the experiment to compare the natural viability of the cell culture. The second pair of samples, both blank and MNP-loaded DCs (Figures 2c and 2d), were exposed to the AMF ($f$= 260 kHz, H= 16 mT) for 30 minutes and, simultaneously, the temperature of the medium was measured as described elsewhere. [19] After the field exposure, cell viability was measured in the four samples using Trypan blue and FACS (annexin-propidium iodide) protocols, as previously described.

2.6. Morphological analysis by scanning electron microscopy (SEM)

We investigated the possible effects of the alternating magnetic fields on cell morphology using scanning electron microscopy (SEM). Images taken of MNP-loaded cells with and without hyperthermia were obtained with a FEI INSPECT F microscope. The samples were prepared by fixing the cells with 2.5% glutaraldehyde in 0.1 M sodium cacodylate and a 3% sucrose solution for 90 min at 37ºC. The dehydration process was conducted by incubating the cells for 5 min with increasing concentrations of methanol and anhydride methanol for 10 minutes. All samples were sputter-coated with gold immediately before observation.

## *3. Results and discussion*

### 3.1. Magnetic nanoparticles

Figure 2 shows a typical TEM image of the colloids used in this work. As a general feature, the magnetic cores possessed cubic-rounded morphology, with some degree of agglomeration, even after strong ultrasonication and extensive dilution. Therefore, the presence of agglomerates in the colloidal suspensions could not be eliminated. The average sizes of the magnetic cores were obtained after counting over 100 nuclei from different images and fitting the resulted histogram with a log-normal function. This shape was chosen after finding a notable fraction of large, cubic-shaped particles in both samples. The resulting diameters were $<d> = 13\pm2$ and $15\pm2$ nm, and the dispersion values were $\sigma = 0.5$ and $0.6$ for samples NH2(MNP)+ and COOH(MNP)-, respectively.

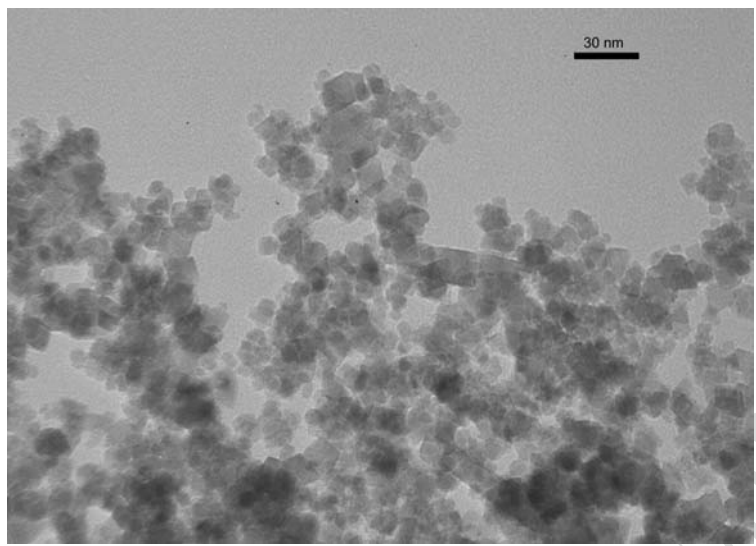

*Figure 2:* TEM image of Fe$_3$O$_4$ nanoparticles with magnetic cores of ⟨d⟩ = 13(2) nm, functionalised with NH$_2^+$ groups at the surface.

The hydrodynamic diameters of the colloidal nanoparticles were extracted from the DLS curves at room temperature. Both NH2(MNP)+ and COOH(MNP)- samples displayed a log-normal profile (Figure 3), with a broad peak at approximately 240 nm and a 'tail' from the contribution of larger particle sizes. The large value of d$_H$ (≈18 times the size of the magnetic core) exemplified the importance of the organic coating for biomedical applications because the particle aggregates that were close to the micrometer range would have likely been difficult to incorporate by the usual endocytosis mechanisms. Because DLS measurements were performed on highly diluted colloids, the observed dispersion of the particle sizes ($\sigma_H$ = 0.40-0.50 values of the fitting log-normal function) reflected the presence of clustered units in the colloid, which was in agreement with the agglomeration observed in the TEM images discussed above.

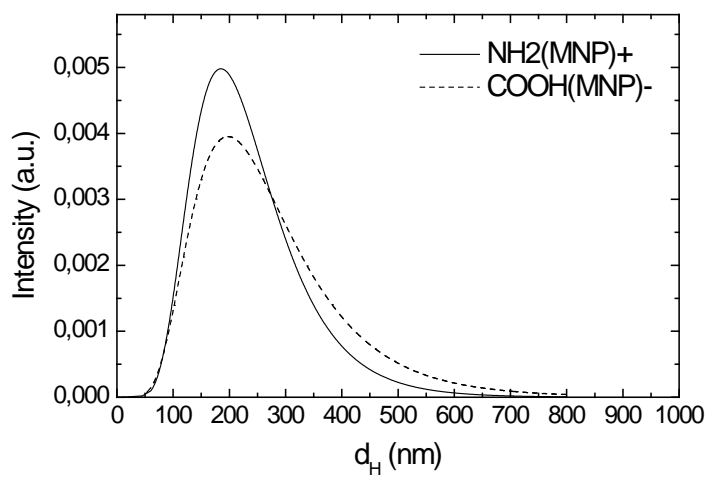

**Figure 3.** DLS results for the NH2(MNP)+ (solid line) and COOH(MNP)- (dashed line) nanoparticles.

The SPA values for the different types of commercial MNPs tested varied from 18 to 36.6 W/g (Figure S1 in the supporting material). The SPA is known to depend strongly on the average particle size, size dispersion, particle interactions (through magnetic dipolar interactions) and the hydrodynamic diameter. [20, 21] There is usually a narrow size window where the SPA is maximised; this window is centred at some value that depends on magnetic parameters such as saturation magnetisation and magnetic anisotropy of the core material. Because the purpose of this study was to induce cell death with a minimum amount of MNPs, the two samples having a) the highest SPA values and b) opposite surface charge were selected for investigation of the optimal cellular uptake. The pair of samples with the highest SPA values and opposite surface charges corresponded to samples NH2(MNP)+ (SPA=36.8 W/g) and COOH(MNP)- (SPA=46 W/g).

Figure 4 shows the magnetisation data M(T) as a function of the temperature of both NH2(MNP)+ and COOH(MNP)- samples, obtained in zero-field cooling (ZFC) and field-cooling (FC) modes. The similar behaviours reflect the similar size and composition of the magnetic cores of both samples. The continuous increase of the M(T) data taken in ZFC mode (Figure 4) suggest that the blocking temperature, $T_B$, was close to the maximum measured temperature. The single-domain nature of the particles was inferred from the absence of the jump in M(T) curves in the ZFC mode, which is expected for the Verwey transition in multidomain particles.[22-24] In agreement with the proximity of the $T_B$ inferred from the M(T) data, the hysteresis loops M(H) performed at 295 K in the lyophilised colloids showed nearly a zero coercive field $H_C$ (see Figure 4), which indicates the presence of single-domain nanoparticles in the superparamagnetic regime. The magnetic parameters extracted from the M(H) curves at low temperatures (Figure S2 of the supporting information) indicated a slightly reduced $M_S$, which is frequently found in $Fe_3O_4$ particles with average size $<d>\gtrsim 10$ nm and attributed to surface spin disorder. [22] The relevant magnetic parameters are summarised in Table I.

**Table I:** Chemical and physical parameters of the nanoparticles used in this work: surface functional groups, core size $d_{CORE}$, hydrodynamic size ($d_H$), distribution ($\sigma_H$), saturation magnetisation ($M_S$), coercive field ($H_C$) at room temperature, and specific power absorption (SPA) values.

| Sample | Core material | Surface charge | $d_{CORE}$ (nm) | $d_H$ (nm) | $\sigma_H$ | $M_S$ (emu/g) | $H_C$ (Oe) | SPA (W/g) |
|---|---|---|---|---|---|---|---|---|
| **NH2(MNP)+** | $Fe_3O_4$ | $NH_2^+$ | 13(2) | 244(2) | 0.48 | 62.3 | 12(5) | 36(6) |
| **COOH(MNP)-** | $Fe_3O_4$ | $COOH^-$ | 15(2) | 217(2) | 0.43 | 68.0 | 12(5) | 46(6) |

In summary, the magnetic characterisation of both colloids used in this work indicated a single-domain structure of the MNP core with superparamagnetic behaviour in agreement with previous results for $Fe_3O_4$ nanoparticles at similar sizes.[22, 24]

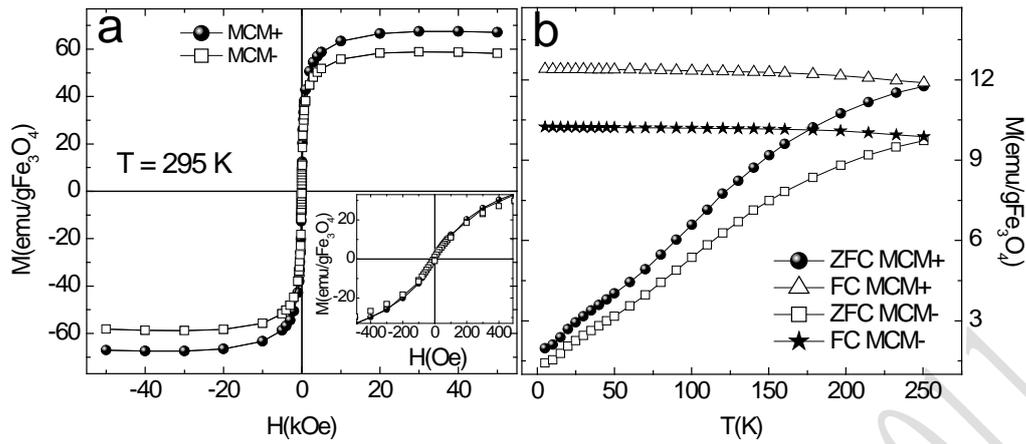

**Figure 4:** Magnetic data from magnetic colloids used in this work. a) Hysteresis loops *M(H) at room temperature. The inset shows the small coercive field; b)ZFC-FC data for the same colloids, showing the single-domain behaviour with blocking temperatures above 250 K.*

### 3.2. Dendritic cell differentiation from PBMCs

Monocytes (CD14+) cultured for 5 days in the presence of GM-CSF and IL-4 showed the characteristic phenotype of immature DCs. As shown in Figure 5, the expression of the main markers characterising immature DCs was confirmed: CD45 (a leukocyte common antigen); CD40 (a common feature of TNF receptor family members, which play an important role in the maturation of DCs) and CD11c (a transmembrane protein present at high levels on most human DCs). Additionally, the absence of specific markers, such as CD14 (expressed by monocytes but not expressed by DCs), CD83 and CD86 (typically absent for immature DCs but present after maturation), was observed. The combination of the above results (both positive and negative) confirmed that the phenotype of the cultured cells corresponded to immature DCs. This result was a requisite for the vectorization of MNPs to neoplastic tissue because the overall strategy relies on the ability of immature DCs to transdifferentiate into endothelial cells when cultured in the presence of angiogenic growth factors [25] or tumour-conditioned media.[26] When DCs are stimulated by antigens and consequently reach a mature state, they lose their multipotency to differentiate into other types of cell.[27, 28] Therefore, giving the immature phenotype sufficient time for the DCs to reach the desired organ is essential for any potential therapeutic application.

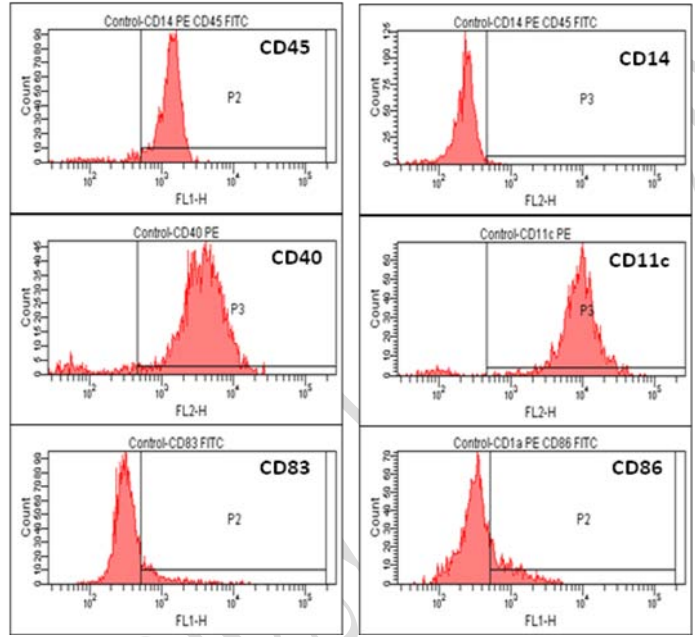

**Figure 5:** *Surface antigen expression measured by FACS for 5-day-cultured cells. The labels in each figure denote the type of FITC or PE-conjugated antibodies used in each case. The resulting phenotypes were immature DCs.*

### 3.3. Viability of DCs cultured with MNPs

The cytotoxicity produced by MNPs was another obstacle that needed to be overcome for biomedical applications. Therefore, the first step to evaluate the actual potential of this strategy was to determine whether DCs could incorporate this type of MNPs without affecting the viability of the cells. In a previous work, Wang et al. [29] studied the uptake of multiwalled carbon nanotubes (CNT) by immature dendritic cells (iDCs) and found that the incorporation of CNTs did not induce phenotypical changes in iDCs and did not affect the normal activation by lipopolysaccharides. Polystyrene particles have also been widely used for this purpose because they are commercially available with different fluorescent markers, surface charges and well-defined sizes (from a few nanometres to a few microns). [30] The use of these types of particles has demonstrated that the cytotoxic effects depend on both the size and surface charge.[30, 31] The potential of biodegradable nanoparticles for antigen delivery using DCs has been demonstrated *in vivo* by Matsuo et al. [32], who reported the inhibition of tumour growth in C57BL/6 mouse experiments.

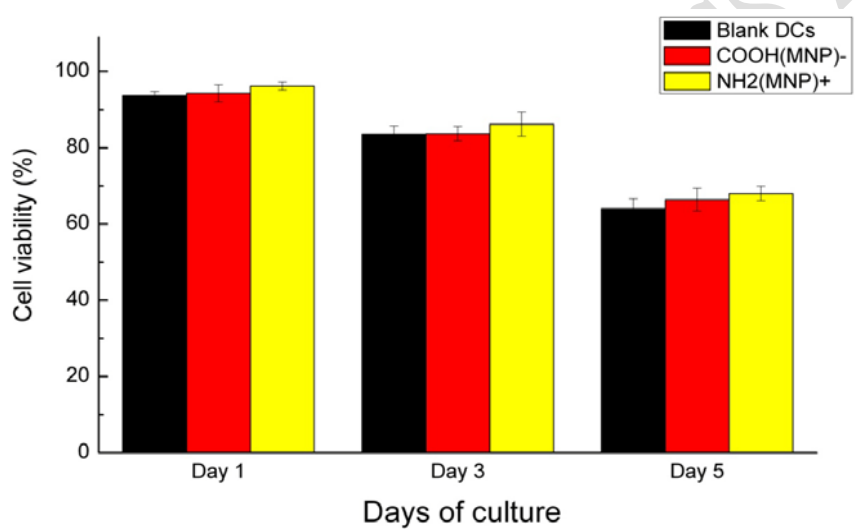

**Figure 6:** *Trypan blue results of cell viability for several five-day experiments with DCs loaded with positively (NH2(MNP)+) and negatively (COOH(MNP)-) charged nanoparticles. The primary DC culture without MNPs showed a natural decrease of viability from approximately 95% to 64 % on the fifth day. The same trend, within experimental error, was displayed by DCs loaded with both types of MNPs. The data and error bars represent the mean and SD of n=3 independent experiments.*

The cytotoxicity of the MNPs used in this work was verified by Trypan blue, flow cytometry and MTT analyses. The obtained results are shown in Figure 6, where the viability of DCs exhibited a decreasing trend from day 1 to 5 of co-culture for both NH2(MNP)+ and COOH(MNP)- in a manner similar to that of the control sample (without MNPs); the viability remained above 65-70% on the fifth day of culture. It is worth mentioning that the DCs used in this work are primary cells; therefore, they have a finite life span in culture, as opposed to immortalised cell lines. Flow cytometry and MTT assays (Figures S3 and S4 in the supporting material) confirmed that the decrease of cell viability after five days was not affected by the incorporation of MNPs.

### 3.4. TEM study of internalisation

Optimisation of the uptake of MNPs by the DCs was required before these cells could be used as carriers to the tumoural vasculature.[33] This, in turn, required the determination of the uptake mechanisms involved for a given type of particle and cell line because these mechanisms would determine the uptake kinetics and intracellular fate of the nanoparticles. Endocytosis has been recognised as the most common passive mechanism for the MNP uptake by different cellular types, including HeLa cells, fibroblasts and DCs. The intracellular localisation of the MNPs was evaluated by TEM images of DCs cultured with MNPs during a 24-h period. In Figure 7, a series of representative TEM images illustrated that the uptake sequence in DCs was a stepwise process in which the MNPs were first attached to the cell surface and subsequently internalised by vesicle-mediated endocytosis. After attaching to the external side of the cell (*Figure 7*a), the invagination of the cell membrane began at the areas of contact (the arrow in *Figure 7*b) to form an endosomal structure that finally progressed to the cell interior with most of its volume occupied by the endocytosed MNPs (*Figure 7*c-d). Although the different stages of the uptake process were present in all samples observed, the fraction of mature endosomes was predominant.

The above results were in agreement with previously reported data [34] showing that monocyte-derived DCs were able to incorporate Fe@C magnetic nanoparticles and that the intracellular distribution of these MNPs was primarily located in the cytoplasmatic compartments.

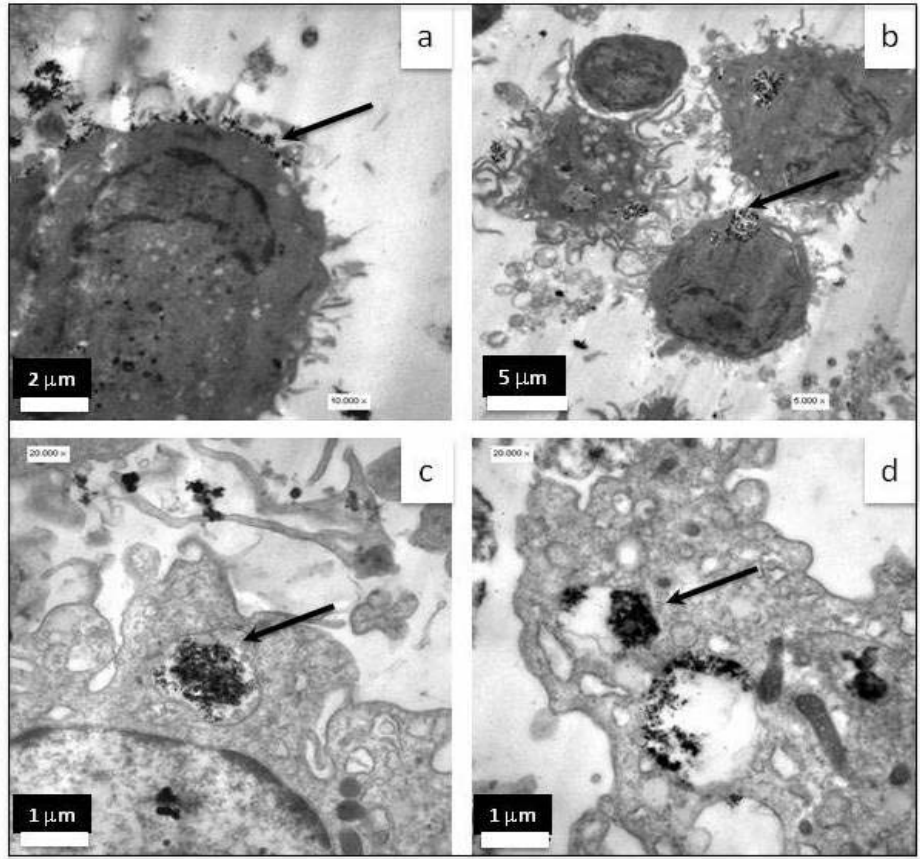

**Figure 7:** TEM images of DCs incorporating COOH(MNP)- nanoparticles. The sequence of the four panels represents a possible path for MNP incorporation: (a) the arrow indicates the presence of MNPs bound to the cell surface. The invagination of the cell membrane can be clearly seen in (b), capturing the attached MNPs. The lower panels (c) and (d) show the completed endosome with large amounts of MNPs inside.

Several studies have also reported the uptake of different complexes, such as latex-type nanospheres, by DCs [35, 36] and the capacity of DCs to incorporate a wide range of antigens and molecules [37, 38]. In many cell lines, the uptake of MNPs occurs in a concentration- and time-dependent manner[39] that depends on some specific features of the MNPs involved. In the absence of surface functionalisation of MNPs with specific biologic ligands, as in the case of the present MNPs, the most probable mechanism involved in MNP uptake is nonspecific endocytosis.

### 3.5. Quantification of uploaded MNPs

Magnetic measurements of the MNP-loaded DCs were performed as a simple way of quantifying the amounts of magnetic material incorporated. Figure 8 shows the low-temperature M(H) hysteresis loops for DCs cultured with COOH(MNP)- and NH2(MNP)+ particles at the same concentration of 50 μg MNP/ml medium. All data refer to $10^6$ cells and a control sample (i.e., $1 \times 10^6$ DCs without MNPs), which was measured at the same temperature to subtract the diamagnetic signal from the DCs at low temperatures (see the lower inset in Figure 8). The similar saturation magnetisation extracted from both M(H) curves revealed that the incorporation of MNPs did not depend on the surface charge (Figure 8). Similar experiments performed on DCs cultured with increasing concentrations (50, 150 and 300 μg MNPs/ml medium) demonstrated that the MNPs were incorporated into DCs in a concentration-dependent manner (see Figure S5 in supporting material). A non-linear increase of the magnetic signal was observed for higher concentrations of MNPs added, which was likely due to excess MNPs remaining in the culture medium after washing of the cultures. At a high MNP concentration, a brownish colour was observed for the cell medium, which indicated that excess magnetic nanoparticles mixed with the medium and could not be totally removed; this influenced the total magnetic signal.

Using the $M_S$ values for the pure ferrofluid and the average volume of the magnetic cores extracted from TEM images, the average number of MNPs per cell was determined to be $1 \times 10^4$ to $7 \times 10^4$ MNPs/cell. This relatively small number of particles was able to induce cellular death when submitted to an AMF, as will be discussed in the next sections.

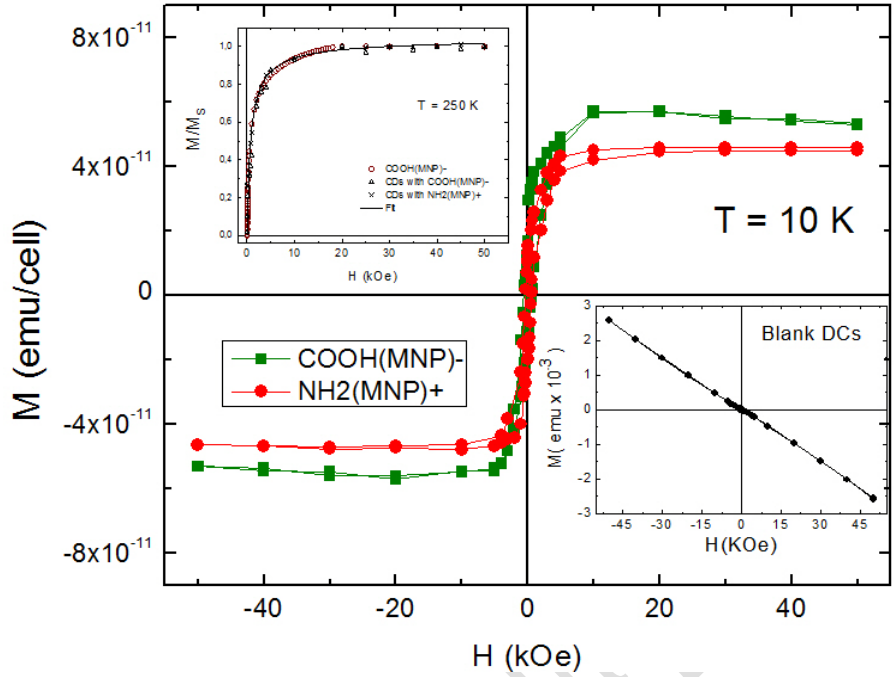

**Figure 8: Main panel:** low-temperature magnetic signal from $1\times10^6$ DCs cultured with COOH(MNP)- (squares) and NH2(MNP)+ (circles) at a concentration of 50 μg MNPs/ml medium. The diamagnetic signal of $1\times10^6$ blank DCs (lower inset) was subtracted in all cases. **Upper inset:** normalised initial magnetisation $M(H)/M_S$ curves at T = 250 K. The scaling of pure ferrofluid (open circles) and both types of MNPs inside the DCs (triangles and crosses) reflects the non-interacting behaviour of the MNPs. The solid line is the corresponding Langevin fit using equation (1).

To study the physical state of aggregation of the MNPs in the intracellular space, the magnetic state of the MNPs at room temperature was analysed in samples of DCs+MNPs that were previously lyophilised. Similar to the pure colloids at room temperature (Figure 4), the particles uploaded by the cells displayed SPM behaviour reflected in the M(H) curves with null coercivity at 250 K (upper inset of Figure 8). The magnetisation of such particles in the SPM state is described by the expression $M=M_S \, L(x)$, where $M_S$ is the saturation magnetisation and $L(x)$ is the Langevin function with $x=\mu H/k_B T$, μ is the magnetic moment of a single MNP, H is the applied field, $k_B$ is the Boltzmann constant and T is the temperature. To take the volume distribution into account, the SPM magnetisation of the particles was better described as a weighted sum of Langevin functions [22].

$$M = \int_0^\infty L\left[\frac{\mu(V)H}{k_B T}\right] f(V)dV \qquad (1)$$

where $f(V)$ represents the distribution function of the particle volume, which is usually assumed as a log-normal function. It should be noted here that this expression assumes that the system is composed of *noninteracting* particles. The fit of the M(H) curves at room temperature using Eq. (1) for a) *as-prepared* ferrofluid and b) magnetically loaded DCs resulted in identical fitted parameters within error, as shown by the scaling of the corresponding M(H) curves in the upper inset of Figure 8. This result indicates that a) the $Fe_3O_4$ phase was not reduced/oxidised in the intracellular medium and that b) the particle size distribution remained essentially unaltered after the incorporation into the DCs (i.e., no selective sizes were incorporated). As discussed previously, the TEM images showed that MNPs were agglomerated inside the endosomal spaces[34]; therefore, it was concluded that the dextran/functionalised shell was sufficiently thick to prevent strong interparticle interactions. This is an important issue for heating experiments *in vitro* because strongly interacting MNPs could result in a decrease of the power absorption. [40]

### 3.6. Power absorption experiments

Once the incorporation of MNPs was verified and the cytotoxicity was evaluated, the effects of AMF were studied in MNP-loaded DCs. The heating efficiency of the magnetic colloids was measured through the absorbed power P divided by the mass $m_{NP}$ of the constituent nanoparticles diluted in a mass $m_{LIQ}$ of the liquid carrier. The expression for the SPA can be written as follows [6]:

$$\Pi = \frac{P}{m_{NP}} = \frac{m_{LIQ}c_{LIQ} + m_{NP}c_{NP}}{m_{NP}}\left(\frac{\Delta T}{\Delta t}\right), \qquad (2)$$

where $c_{LIQ}$ and $c_{NP}$ are the specific heat capacities of the liquid carrier. Because the concentrations of MNPs are usually in the range of 1 % wt., we can approximate [21] $m_{LIQ}\,c_{LIQ} + m_{NP}\,c_{NP} \approx m_{LIQ}\,c_{LIQ}$ and (2) can be written as follows:

$$\Pi = \frac{c_{LIQ}\delta_{LIQ}}{\phi}\left(\frac{\Delta T}{\Delta t}\right), \qquad (3)$$

where $\delta_{LIQ}$ and $\phi$ are the density of the liquid and the weight concentration of the MNPs in the colloid, respectively. The conversion from magnetic field energy into heat by the absorbing elements (i.e., the magnetic nanoparticles) at the frequencies of the experiment was due to the Néel relaxation process. Many recent reports have shown the details of this mechanism, which involves heat transference from nanoscopic agents (MNPs) to the macroscopic scale of the colloid. The common finding is that there is a strong dependence of SPA on particle size, with a narrow peak of absorption near the single- to multi-domain transition.[41, 42] As this represents a heat transference mechanism, an *adiabatic* experiment was needed to obtain actual SPA values for the given material by measuring the exact energy released as the temperature increased within the sample. Because the heating of a macroscopic amount of the sample took $10^2$ to $10^3$ seconds, high insulation was needed to provide adiabatic conditions during the whole experiment. Complete insulation is seldom attained in hyperthermia experiments [43], as can be inferred from the measurable decrease of the temperature when the field was turned off. This condition hindered the accurate determination of the final temperature that could be attained with a given set of nanoparticles. However, the criterion for choosing a representative SPA value was related to the maximum slope of the T vs. time curves, which occurred during the first seconds of the experiments; thus, it can be ensured that the system behaved adiabatically because the amount of heat lost during this short interval could be neglected. Conversely, *in vitro* and *in vivo* power-absorption experiments were performed under strong heat-exchanging conditions, and therefore, it was neither possible nor interesting to quantify the power absorbed by the MNPs or the cells. Although the measured parameter in the hyperthermia experiments was the temperature increase of the cell culture, the resulting cellular damage had to be related to the actual intracellular temperatures attained. However, to the best of our knowledge, there are no detailed studies on the dynamics of heat conduction from the intracellular medium through the cell membrane.

For each power absorption experiment, the DC cultures were always conducted in parallel wells (i.e., two wells each for blank and MNP-loaded DCs) to compare the viability between cells that were submitted to AMF and those that were not. The results of these experiments (Figure 9) indicated 70-95% viability in the blank samples and loaded DCs with no exposure to the magnetic field (figures 9a and 9b) and in the MNP-unloaded DCs subjected to AMF for 30 min (Figure 9c). Conversely, the application of the magnetic field to MNP-charged DCs resulted in a sharp decrease of cell viability to 2-5 % (Figure 9d), demonstrating the efficacy of an AMF for killing the targeted cells. Data from the Trypan blue viability analysis (not shown) confirmed these results regarding the viability of targeted DCs with and without the application of AMF. A complementary series of experiments conducted using DCs loaded with NH2(MNP)+ nanoparticles (Figure S6 of the supporting information) showed the same effect of the AMF on MNP-loaded DCs.

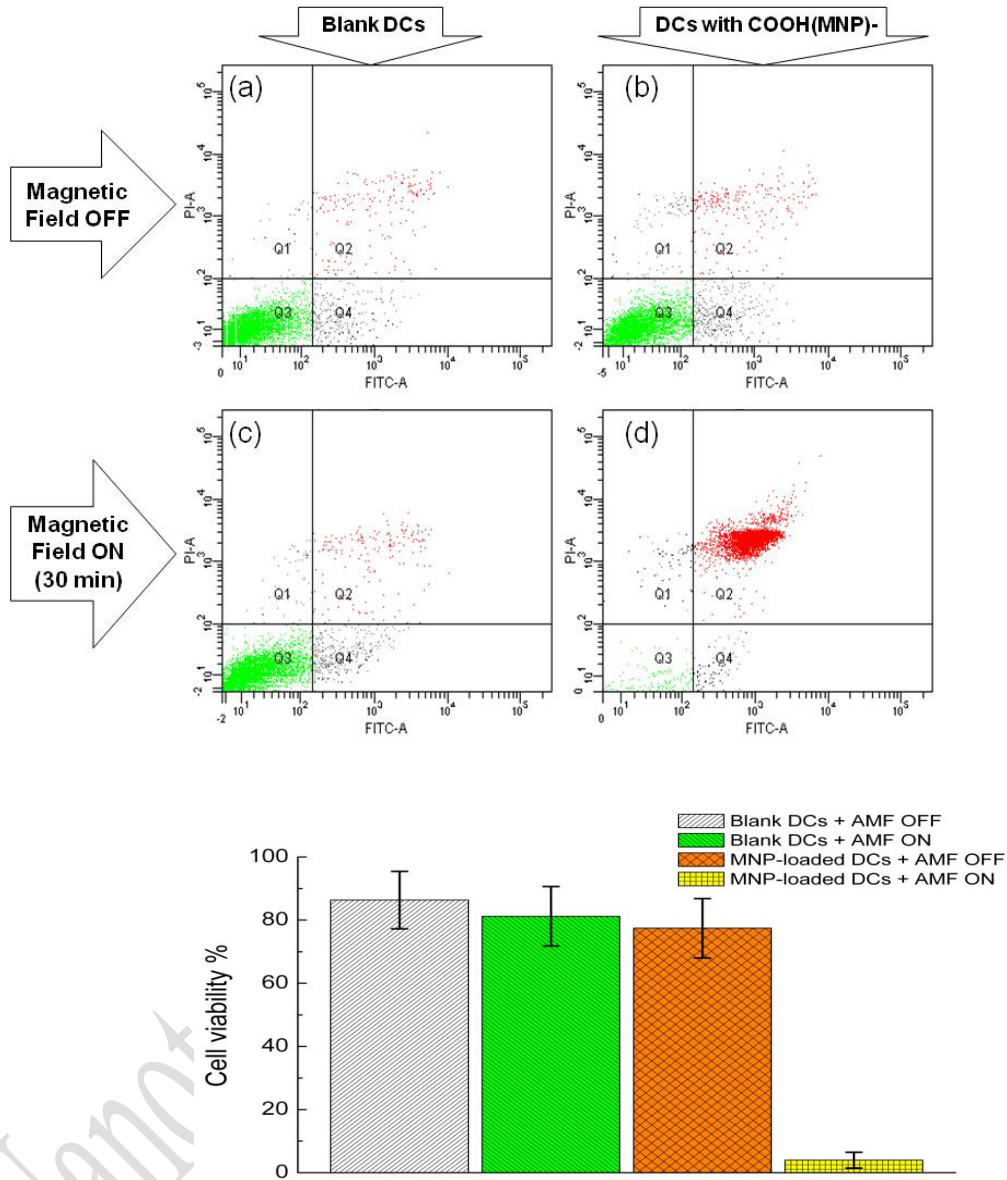

**Figure 9:** *A typical experimental result for the '2x2' experiments schematised in Figure 1. Viability from FACS data for: (a) non-loaded DCs with magnetic field OFF, (b) DCs with COOH(MNP)- particles with magnetic field OFF, (c) non-loaded DCs after application of 30 min of AMF and (d) DCs loaded with COOH(MNP)- particles after application of 30 min of AMF. The green dots at quadrant Q3 indicate the viable cells in each experiment.* **Lower panel:** *Same data and error bars representing the mean and SD (n=3) of the '2x2' experiments.*

Interestingly, the temperature of the cell medium measured by the macroscopic optic sensor remained within 29-31 ºC (see Figure 10) (i.e., a few degrees above the starting temperature of 26 ºC) but well below the values of 43-46 ºC needed to trigger temperature-induced apoptosis by hyperthermia. The absence of a macroscopic temperature increase during the cell death observed for samples d) and f) raised the question of a possible intracellular mechanism that differed from the temperature increase of the intracellular medium.

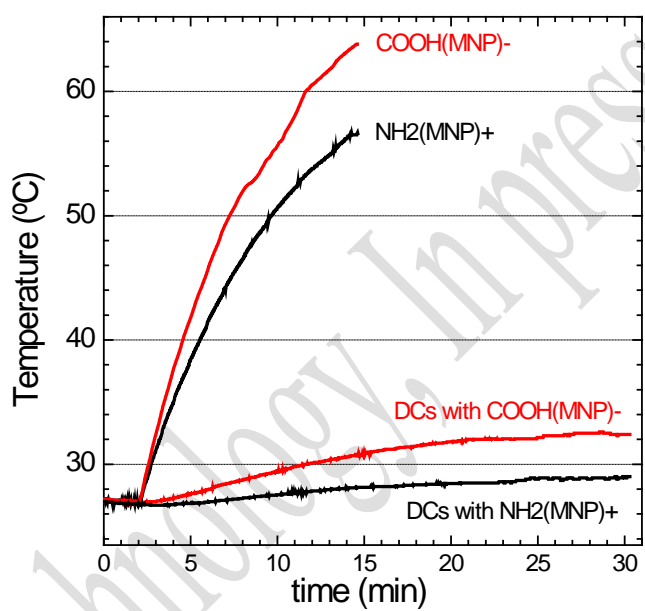

**Figure 10:** Heating curves for pure colloids NH2(MNP)+ and COOH(MNP)-, during the application of alternating magnetic fields (H = 160 Oe, $f$ = 260 kHz). The same experiment for DCs loaded with NH2(MNP)+ and COOH(MNP)- (bottom curves) produced a much smaller temperature increase to 29-32 ºC after 30 min of exposure to the AMF.

In a series of comprehensive experiments using $CoFe_2O_4$ and $\gamma\text{-}Fe_2O_3$ magnetic MNPs, Fortin *et al*. showed that heating can be achieved with a concentration of 25-30 pg at the intracellular level.[44] This concentration was larger than that observed in the experiments herein, in which a maximum concentration of 5 pg MNP/cell was obtained after incubation in medium with 300 μg MNP/ml during a 12 h period (as shown in Figure S5 of the supporting material); the main reason for these differences could be related to the different affinities for MNPs among different cell lines. As observed in Figure 10, the total amount of MNP taken up after co-culture was not sufficient to raise the temperature of the whole cell culture when submitted to an AMF in the thermally insulated system. This was consistent with the low average concentration of MNPs (2 μg MNPs/ml medium) in the experiments, as calculated from the $10^6$ DCs containing an average of 1.07 pg MNPs each and dispersed in 0.5 ml of medium. If the calculation for the MNP concentration was restricted to the intracellular medium with the internalised MNP amount of 1.07 pg/cell and considering an average cell volume (from SEM images) of $V_{cell}$=500 μm$^3$, then the resulting intracellular MNP concentration would be ≈2 mg/ml, which was not sufficient to raise the temperature of the intracellular medium with the present experimental conditions (H= 160 mT, $f$= 260 kHz). It was concluded that the temperature increase of the cell culture was not a relevant parameter for induction of cell death by an AMF in MNP-loaded DCs.

Apoptosis or cellular programmed death is a particular mechanism in which a cell follows a programmed sequence of events to "prepare" for its death with minimal disturbance of the whole population. This phenomenon seems to be present in a wide range of organisms from primitive to higher eukaryotes. An early event that is considered a marker of this process is the exposition of phosphatidylserine on the external surface of the plasma membrane. The FACS results of Figure 9 indicate that incubation of DCs with annexin (which reveals the translocation of phosphatidylserine) and propidium iodide (which reveals damage in the plasma membrane) are both positive only for magnetically charged DCs after the application of the AMF (Figure 9d and Figure S6 d in the supporting material).

It is important to stress that the experimental protocols were designed to perform the viability analysis immediately after exposure to the AMF. In spite of the short time elapsed between the AMF exposure and FACS measurement (approximately 2 hours), a clear shift of the cell populations along both axes (samples d and f) was observed, which was evidence of late apoptotic or necrotic cells.

The SEM images obtained from the cell cultures (see Figure 11) showed a clear effect on the membrane integrity after AMF experiments in the DCs loaded with nanoparticles, whereas the cells without loaded MNPs retained the complex membrane structure of functional DCs. The short time elapsed and the large damage observed at the membrane level support the hypothesis of a necrosis-like process. The large channels observed in Figure 11(b) suggest that after exposure to the AMF, the cells lose all membrane activity and become permeable. The origin of such a dramatic effect attained by a relatively small amount of MNPs ($\approx$1 pg MNPs/cell) is not yet clear, but the short time involved (30 min of AMF application and a one-hour interval before cells were fixed for SEM images) suggests that the process is of a physical nature as opposed to a programmed loss of metabolic activity followed by cell death. In addition, the release of the power absorbed by the MNPs did not yield *a macroscopic* temperature increase when the temperature was measured in the whole culture medium by a macroscopic probe. The possibility remains, however, that the temperature of the intracellular medium increased during the AMF application, resulting in metabolic and/or cytoskeleton damage. This hypothesis must be tested using a yet-undeveloped microscopic, intracellular temperature probe, and work along this direction is currently being performed.

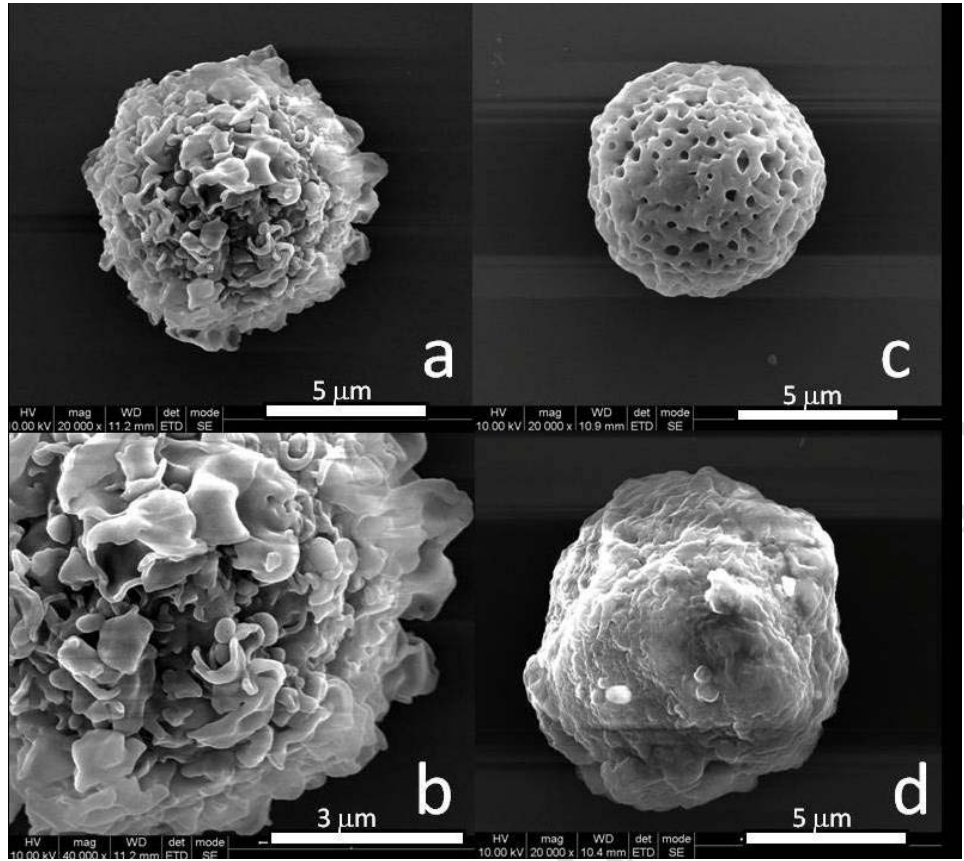

**Figure 11:** *SEM images of DCs after the application of the AMF for 30 min: a) and b) cells not loaded with COOH(MNP)-; c) and d) DCs loaded with MNPs. Note the evident loss of membrane structure and 'shrinking' of the cells in c) and d). Note also in c) the large channels opened in the membrane.*

In spite of the experimental difficulty for measuring intracellular temperatures, some theoretical calculations were performed to evaluate the capability of the number of MNPs incorporated to release sufficient energy to damage the subcellular structure. From the average particle diameter of the MNPs measured by the TEM images, the number of particles within a single cell was estimated to be $\approx 1 \times 10^4$. Assuming the same power absorption as the pure ferrofluid (36-46 W/g, see **Table I**), this amount of MNPs could release a total power of $\approx 2 \times 10^{-10}$ W to the nearest organelles. If a short release time of 10 s was considered, the total energy available for local damage could be on the order of $1 \times 10^{10}$ eV. Compared to typical binding energies per atom in organic molecules ($E_{C-N}$=308 eV, $E_{C-C}$=348 eV, $E_{C=C}$=614 eV), the previous estimation indicates that there should be sufficient energy available to induce bond breakage in the organic structures surrounding the MNPs.

The confinement of these particles within endosomal structures (as observed from the TEM images) drastically reduced their contact with intracellular structures. Therefore, it seems plausible that the MNPs agglomerates observed inside the endosomes would be able to disrupt the endosomic membrane during AMF application. Endosomes and lysosomes are acidic vesicles involved in the endocytic pathway, and the lysosomes are the terminal degradation compartment of the phagocytosed material. Thus, the consequences of the local damage to the lysosomal membrane (i.e., the release of the lysosomal content into the cytoplasm) could induce changes in the intracellular medium that could lead to cell death.

Although a complete analysis of the actual mechanism at the cellular level would require a study of a variety of markers, such as proteases, caspases, and cytokines, the large fraction of dead cells observed in such a short time suggests that a necrosis-like process was the main cause of cell death during magnetic field application. The precise nature of this mechanism is of crucial importance for future application protocols because necrosis must always be avoided *in vivo* due to the lethal consequences to the surrounding tissues. If the energy release is the cause of cellular damage, the application time and power could be regulated to a value at which only apoptosis is triggered.

The combined results of a) low cytotoxicity of the magnetic MNPs and b) cell death after the application of an AMF suggest that MNP-loaded DCs could be used as magnetic carriers to tumoural regions, where cellular death induced by an external magnetic field could be used as a therapy against tumour growth.

## *4. Conclusions*

Based on the described ability of DCs to be loaded with magnetic nanoparticles, they appear to be a potentially powerful 'Trojan horse' system for MNP delivery to specifically targeted sites. The results indicate that the MNPs had negligible cytotoxic effects and that the functionality of the DCs was maintained, which supports the feasibility of this approach. Therefore, the demonstration of AMF-induced cell death makes the selective killing of MNP-loaded cells feasible. The detailed intracellular mechanism has yet to be determined; however, preliminary results suggest an initial disruption of the endosome/lysosome membrane by the MNPs that resulted in the release of their content with subsequent cell death.

In combination with magnetic nanoparticles and hyperthermia treatments, DCs may become a useful tool for many biomedical applications. The ability of DCs to transdifferentiate into endothelial-like cells under pro-angiogenic conditions and interact with the tumoural vasculature has been previously reported. [45] Therefore, DCs loaded with MNPs *ex vivo* could be vectorized or injected *in situ* at tumour sites to the pro-angiogenic microenvironment present in the tumour area. DCs might differentiate to endothelial-like cells and integrate into the new vasculature. Death of the MNP-loaded cells and, therefore, the tumour blood supply through the exposure to AMFs could prevent tumour vessel network formation and constitute a highly selective treatment for cancer patients.

## 5. Acknowledgements

This work was supported by the Ministerio de Ciencia e Innovación (project MAT2008-02764) and the Diputación General de Aragón (DGA) and IBERCAJA. We are grateful to Javier Godino [Hospital Clínico Universitario Lozano Blesa, Instituto Aragonés de Ciencias de la Salud (I+CS), Zaragoza] for his assistance with flow cytometer measurements. The help of Iñigo Echaniz (technical support) is also gratefully recognised.

*Financial & competing interests/ disclosure*

G.F.G. and M.R.I., along with the University of Zaragoza, have filed patents related to the technology and intellectual property related to the magnetic field applicators reported herein. G.F.G. and M.R.I. have equity in nB Nanoscale Biomagnetics S.L. The other authors declare that they do not have any affiliations that would lead to a conflict of interest.